\documentclass{article}
\usepackage{graphicx}

\def\Journal#1#2#3#4{{#1} {\bf #2}, #3 (#4)}

\def\NPA{{\em Nucl. Phys.} A}

\def\PLB{{\em Phys. Lett.} B}

\def\PRL{\em Phys. Rev. Lett.}

\def\PRD{{\em Phys. Rev.} D}
\def\PRC{{\em Phys. Rev.} C}

\def\ZPA{{\em Z. Phys.} A}

\begin{document}

\title{
HADRONS IN THE NUCLEAR MEDIUM \\ -- INTRODUCTION AND
OVERVIEW\footnote{work supported by BMBF and DFG}\footnote{Based on results of work with E. Bratkovskaya, W. Cassing, M.
Effenberger, H. Lenske, S. Leupold, W. Peters, M. Post and T. Weidmann.}}

\author{U. MOSEL\footnote{mosel@theo.physik.uni-giessen.de}\\
Institut fuer Theoretische Physik, Universitaet Giessen\\
D-35392 Giessen, Germany}

\maketitle

\abstract{In this talk I first discuss predictions based on QCD sum
rules and on hadronic models for the properties of vector mesons in the
nuclear medium. I then describe possible experimental signatures and
show detailed predictions for dilepton invariant mass spectra with a
special emphasis on nuclear reactions involving elementary incoming
beams and nuclear targets. I will in particular illustrate that the
sensitivity of pion and photon induced reactions to in-medium effects
is comparable to that of heavy-ion reactions.}

\section{Introduction}
The investigation of in-medium properties of hadrons has
found widespread interest during the last decade. This interest was
triggered by two aspects.

The first aspect was a QCD sum-rule based prediction by
Hatsuda and Lee\cite{HL} in 1992 that the masses of vector mesons
should drop dramatically as a function of nuclear density. It was widely
felt that an
experimental verification of this prediction would establish a
long-sought direct link between quark degrees of freedom and nuclear
hadronic interactions. In the same category fall the predictions of
Brown and Rho that argued for a general scaling for hadron masses with
density\cite{Brown-Rho}.

The second aspect is that even in ultrarelativistic heavy-ion reactions,
searching for
observables of a quark-gluon plasma phase of nuclear matter, inevitably
also many relatively low-energy ($\sqrt{s} \approx 2 - 4 \:
\mbox{GeV}$) final state interactions take place. These interactions
involve collisions between many mesons for which the
cross-sections and meson self-energies in the nuclear medium are not
known, but may influence the interpretation of the experimental
results.

Hadron properties in medium involve masses, widths and coupling
strengths of these hadrons. In lowest order in the nuclear density all
of these are linked by the $t \rho$ approximation that assumes that the
interaction of a hadron with many nucleons is simply given by the
elementary $t$-matrix of the hadron-nucleon interaction multiplied with
the nuclear density $\rho$. For vector mesons this approximation reads
\begin{equation}\label{Vself}
\Pi_{\rm V} = - 4 \pi f_{\rm VN} (0) \rho
\end{equation}
where $f_{\rm VN}$ is the forward-scattering amplitude of the vector
meson (V) nucleon (N) interaction. Approximation (\ref{Vself}) is good
for low densities ($\Pi_{\rm V}$ is linear in $\rho$) and/or large
relative momenta where the vector meson `sees' only one nucleon at a
time. Relation (\ref{Vself}) also neglects the Fermi-motion of the
nucleons although this could easily be included.

Simple collision theory\cite{EJ,Kon} then gives the shift of mass and
width of a meson in nuclear matter as
\begin{eqnarray}\label{masswidth}
  \delta m_{\rm V} &=& - \gamma v \sigma_{\rm VN} \eta \rho \nonumber \\
  \delta \Gamma_{\rm V} &=& \gamma v \sigma_{\rm VN} \rho ~.
\end{eqnarray}
Here, according to the optical theorem, $\eta$ is given by the ratio of
real to imaginary part of the forward scattering amplitude
\begin{equation}
  \eta = \frac{\Re{f_{\rm VN}(0)}}{\Im{f_{\rm VN}(0)}} ~.
\end{equation}

The expressions (\ref{masswidth}) are interesting since an
experimental observation of these mass- and width-changes could give
valuable information on the free cross sections $\sigma_{\rm VN}$ which
may not be available otherwise. The more fundamental question, however, is if
there is more to in-medium properties than just the simple collisional
broadening predictions of (\ref{masswidth}).

\section{Fundamentals of Dilepton Production}

From QED it is well known that vacuum polarization, i.e. the virtual
excitation of electron-positron pairs, can dress the photon. Because
the quarks are charged, also quark-antiquark loops can dress the photon.
These virtual
quark-antiquark pairs have to carry the quantum numbers of the photon,
i.e.\ $J^\pi = 1^-$. The $q \bar q$ pairs can thus be viewed as vector
mesons which have the same quantum numbers; this is the basis of
Vector Meson Dominance (VMD).

The vacuum polarization tensor is then, in complete analogy to QED,
given by
\begin{equation}\label{jjcorr}
  \Pi^{\mu \nu} = \int d^4x \, e^{i q x}
  \langle 0 | T[j^\mu (x) j^\nu (0)] | 0 \rangle
  = \left( g^{\mu \nu} - \frac{q^\mu q^\nu}{q^2} \right) \Pi(q^2)
\end{equation}
where $T$ is the time ordering operator.
Here the tensor structure has been exhibited
explicitly. This so-called current-current correlator contains the
currents $j^\mu$ with the correct charges of the vector mesons in
question. Simple VMD\cite{Sak} relates these currents to the vector
meson fields
\begin{equation}\label{VMD}
  j^\mu(x) = \frac{{m_{\rm V}^0}^2}{g_{\rm V}} V^\mu(x)~.
\end{equation}
Using this equation one immediately sees that the current-current
correlator (\ref{jjcorr}) is nothing else but the vector meson
propagator $D_{\rm V}$
\begin{equation}\label{Vprop}
  \Pi(q^2) = \left( \frac{{m_{\rm V}^0}^2}{g_{\rm V}} \right)^2
             D_{\rm V} (q^2) ~.
\end{equation}
The scalar part of the vector meson propagator is given by
\begin{equation}\label{Vprop1}
D_{\rm V}(q^2) = \frac{1}{q^2 - {m_{\rm V}^0}^2 - \Pi_{\rm V}(q^2)}~.
\end{equation}
Here $\Pi_{\rm V}$ is the selfenergy of the vector meson.

For the free $\rho$ meson information about $\Pi(q^2)$ can be obtained
from hadron production in $e^+ e^-$ annihilation reactions\cite{PS}
\begin{equation}\label{hadprod}
 R(s) = \frac{\sigma (e^+ e^- \rightarrow \mbox{hadrons})}
             {\sigma(e^+ e- \rightarrow \mu^+ \mu^-)}
      = - \frac{12 \pi}{s} \Im \Pi(s) ~
\end{equation}
with $s = q^2$. This determines the imaginary part of $\Pi$ and,
invoking vector meson dominance, also of $\Pi_{\rm V}$. The data (see,
e.g. Fig.\ 18.8 in\cite{PS}, or Fig.\ 1 in\cite{KW}) clearly show at
small $\sqrt{s}$ the vector meson peaks, followed by a flat plateau
starting at $\sqrt{s} \approx 1.5 \: \mbox{GeV}$
described by perturbative QCD.

In order to get the in-medium properties of the vector mesons, i.e.\
their selfenergy $\Pi_{\rm V}$, we now have two ways to proceed: We
can, \emph{first}, try to determine the current-current correlator by
using QCD sum rules\cite{HL}; from this correlator we can then
determine the self-energy of the vector meson following
eqs.\ (\ref{Vprop}),(\ref{Vprop1}). The \emph{second} approach consists
in setting up a hadronic model and calculating the selfenergy of the
vector meson by simply dressing its propagators with appropriate
hadronic loops. In the following sections I will discuss both of these
approaches.

\subsection{QCD sum rules and in-medium masses}

The QCD sum rule for the current-current correlator is obtained by
evaluating the function $R(s)$, and thus $\Im \Pi(s)$ (see
(\ref{hadprod})), in a hadronic model on one hand and in a QCD-based
model on the other. The latter, QCD based, calculation uses the fact
that the current-current correlator (\ref{jjcorr}) can be Taylor
expanded in the space-time distance $x$ for small space-like distances
between $x$ and $0$; this is nothing else than the Operator Product
Expansion (OPE)\cite{PS}. In this way we obtain for the free meson
\begin{equation}
R^{\rm OPE}(M^2) = \frac{1}{8 \pi^2} \left(1 + \frac{\alpha_{\rm
S}}{\pi} \right) + \frac{1}{M^4} m_{\rm q} \langle \bar{q} q \rangle +
\frac{1}{24 M^4} \langle \frac{\alpha_{\rm S}}{\pi} G^2 \rangle -
\frac{56}{81 M^6} \pi \alpha_{\rm S} \kappa \langle \bar{q} q \rangle^2
~.
\end{equation}
Here $M$ denotes the so-called Borel mass. The expectation values
appearing here are the quark- and gluon-condensates. The last term here
contains the mean field approximation
$
\langle (\bar{q} q)^2 \rangle = \kappa \langle \bar{q} q \rangle^2
$.

The other representation of $R$ in the space-like region can be
obtained by analytically continuing $\Im \Pi(s)$ from the time-like to
the space-like region by means of a twice subtracted dispersion
relation. This finally gives
\begin{equation}
R^{\rm HAD} (M^2) = \frac{\Re{\Pi^{\rm HAD}(0)}}{M^2}
 - \frac{1}{\pi M^2} \int_0^\infty
ds \, \Im \Pi^{\rm HAD}(s) \frac{s}{s^2 + \epsilon^2}
\exp{-s/M^2}~.
\end{equation}
Here $\Pi^{\rm HAD}$ represents a phenomenological hadronic spectral
function. Since for the vector mesons this spectral
function is dominated by resonances in the low-energy part it
is usually parametrized in terms of a resonance part
with parameters such as strength, mass and width  with a
connection to the QCD perturbative result for the quark structure for
the current-current correlator at higher energies (for details
see Leupold et al.\cite{Leupoldbar} in these
proceedings and refs.\cite{Leup1,Leup2,Lee}).

The QCD sum rule is then obtained by setting
\begin{equation}
R^{\rm OPE}(M^2) = R^{\rm HAD}(M^2)~.
\end{equation}
Knowing the lhs of this equation then allows one to determine the
parameters in the spectral function appearing in $R^{\rm HAD}$ on the
rhs. If the vector meson moves in the nuclear medium, then $R$ depends also
on its momentum. However, detailed studies\cite{Leup2,Lee} find only a very
weak momentum dependence.

The first applications\cite{HL} of the QCDSR have used a simplified
spectral function, represented by a $\delta$-function at the meson mass
and a perturbative QCD continuum. Such an analysis gives a value for the free
meson mass that agrees with experiment. On this basis the QCDSR has been
applied to the
prediction of in-medium masses of vector mesons by making the
condensates density-dependent (for details see\cite{HL,Leup1,Leup2}).
This then leads to a lowering of the vector meson mass in nuclear
matter.

This analysis has recently been repeated with a spectral function that
uses a Breit-Wigner parametrization with finite width. In this
study\cite{Leup1} it turns out that QCD sum rules are compatible with a
wide range of masses and widths (see also ref.\cite{KW}). Only if the
width is -- artificially -- kept zero, then the mass of the vector meson
has to drop with
nuclear density\cite{HL}. However, also the opposite scenario, i.e. a
significant broadening of the meson at nearly constant pole position,
is compatible with the QCDSR. 

\subsection{Hadronic models}

Hadronic models for the in-medium properties of hadrons start from
known interactions between the hadrons and the nucleons. In principle,
these then allow one to calculate the forward scattering amplitude
$f_{\rm VN}$ for vector meson interactions. Many such
models have been developed over the last few
years\cite{KW,HFN,AK,FP,Rapp}.

The model of Friman and Pirner\cite{FP} was taken up by Peters et
al.\cite{Peters} who also included
$s$-wave nucleon resonances. It turned out that in this analysis the
$D_{13}\: N(1520)$ resonance plays an overwhelming role. This resonance
has a significant $\rho$ decay branch of about 20 \%. Since at the pole
energy of 1520 MeV the $\rho$ decay channel is not yet energetically
open this decay can
only take place through the tails of the mass distributions of
resonance and meson. The relatively large relative decay branch then
translates into a very strong $N^* N \rho$ coupling constant (see
also\cite{Rapp,PE}).

\begin{figure}
  \centerline{\includegraphics[height=5cm]{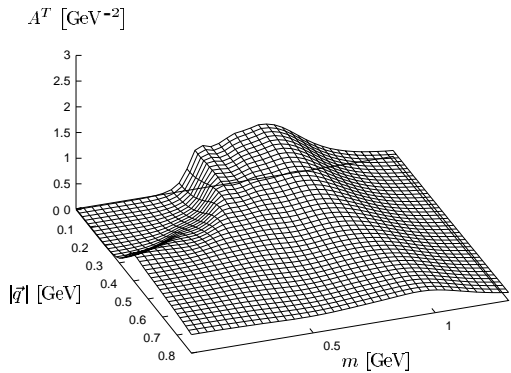}}
  \caption{Spectral function of transverse $\rho$ mesons as a function of
  their three-momentum and invariant mass (from \protect\cite{Peters}).}
  \label{Figrhot}
\end{figure}

The main result of this $N^* h$ model for the $\rho$ spectral
function is a considerable broadening for the latter. This
is primarily so for the transverse vector mesons (see Fig. \ref{Figrhot}),
whereas the longitudinal degree of freedom gets only a little broader with
only a slight change of strength downwards\cite{Peters}.

The results shown in Fig.\ \ref{Figrhot} actually go beyond the simple "$t\rho$"
approximation discussed earlier (see (\ref{Vself})) in that they
contain higher order density effects: a lowering of the $\rho$ meson
strength leads to a strong
increase of the $N(1520)$ $\rho$-decay width which in turn affects the
spectral function. The result is the very broad, rather featureless spectral function for the
transverse $\rho$ shown in Fig.\ \ref{Figrhot}.

\section{Experimental Observables}
In this section I will now discuss various possibilities to verify
experimentally the predicted changes of the $\rho$ meson properties in
medium.

\subsection{Heavy-Ion Reactions}

Early work\cite{Wolf,XK} on an experimental verification of the
predicted broadening of the $\rho$ meson spectral function has
concentrated on the dilepton spectra measured at relativistic energies
(about 1 -- 4 A GeV) at the BEVALAC, whereas more recently many
analyses have been performed for the CERES and HELIOS data obtained at
ultrarelativistic energies (150 -- 200 A GeV) at the SPS. In such collisions
nuclear densities of about 2 - 3 $\rho_0$ can
already be reached in the relativistic domain; in the ultrarelativistic
energy range baryon densities of up to $10 \rho_0$ are predicted (for a
recent review see\cite{BC}). Since the selfenergies of produced vector
mesons are at least proportional to the density $\rho$ (see
(\ref{Vself})) heavy-ion reactions seem to offer a natural
enhancement factor for any in-medium changes.

The CERES data\cite{CERES} indeed seem to confirm this expectation.
The present situation is -- independent of
the special model used for the description of the data -- that agreement
with the measured dilepton mass spectrum in the mass range between
about 300 and 700 MeV for the 200 A GeV $S + Au$ and $S + W$ reactions
can only be obtained if $\rho$-meson strength is shifted downwards (for
a more detailed discussion see\cite{BC,QM}) (for the recently measured 158
A GeV $Pb + Au$ reaction the situation is not so clear; here the
calculations employing `free' hadron properties lie at the lower end of
the experimental error bars\cite{BC}).

However, all the
predictions are based on equilibrium models in which the properties of
a $\rho$ meson embedded in nuclear matter with infinite space-time
extensions are calculated. An ultrarelativistic heavy-ion collision is
certainly far away from this idealized scenario. In addition, heavy-ion
collisions necessarily average over the polarization degrees of freedom.
The two physically
quite different scenarios, broadening the spectral
function or shifting simply the $\rho$ meson mass downwards while keeping
its free width, thus lead to indistinguishable observable consequences
in such collisions.
This can be understood by observing that even in an ultrarelativistic
heavy-ion collision, in which very high baryonic densities are reached,
a large part of the observed dileptons is actually produced at rather
low densities (see Fig. 3 in\cite{Cass}). 

\section{$\pi + A$ Reactions}

Motivated by this observation we have performed calculations of the
dilepton invariant mass spectra in $\pi^-$ induced reactions on
nuclei\cite{Weidmann}; the experimental study of such reactions will be
possible in the near future at GSI. The calculations are based on a
semiclassical transport theory, the so-called Coupled Channel BUU
method (for details see\cite{Teis}) in which the nucleons, their
resonances up to 2 GeV mass and the relevant mesons are propagated from
the initial contact of projectile and target until the final stage of
the collision.  This method allows one to
describe strongly-coupled, inclusive processes without any a-priori
assumption on the equilibrium or preequilibrium nature of the process.

\begin{figure}
  \centerline{\includegraphics[height=8cm]{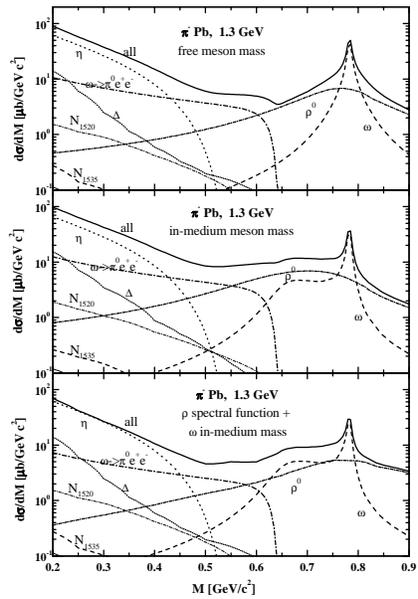}}
  \caption{Invariant mass yield of dileptons produced in pion-induced
  reactions at 1.3 GeV on various nuclei. The topmost part shows results
  of calculations employing free hadron properties, the results shown
  in the middle part
  are based on simple mass-shifts\protect\cite{HL}, and the curves
  in the lowest part show results of a calculation using the spectral function from
  \protect\cite{Peters} for the $\rho$ meson and a
  mass-shift and collisional broadening for the $\omega$ meson (from
  \protect\cite{Weidmann}).} \label{pidilept}
\end{figure}

In these reactions the dominant emission channels are the same as in
ultrarelativistic heavy-ion collisions; this can be seen in Fig.
\ref{pidilept}
where I show the results for the dilepton spectra produced by
bombarding Pb nuclei with 1.3 GeV pions. 

In the top picture in Fig. \ref{pidilept} I show the results of a
calculation that assumes free hadronic properties for all radiation
sources. The middle picture shows the results of a calculation with
lowered in medium masses plus collisional broadening of the vector
mesons, calculated dynamically according to (\ref{masswidth}), and the
bottom figure shows the results of a calculation employing the $\rho$
spectral function calculated in the resonance-hole model discussed
above\cite{Peters}. In the lower 2 pictures the most relevant change
takes place for the $\rho$, which is significantly widened, and the
$\omega$, which develops a shoulder on its low mass tail. The latter is
primarily due to the nuclear density profile: $\omega$'s are produced
at various densities from $\rho_0$ down to 0. 

The dilepton yield in the range of about 600 - 700 MeV is increased by
the in-medium effects by up to a factor of 2.5, comparable to
the ultrarelativistic heavy-ion collisions discussed
above. In particular the $\omega$ shoulder should clearly be visible.

\section{Photonuclear Reactions}

Pion induced reactions have the disadvantage that the projectile
already experiences strong initial state interactions so that many
produced vector mesons are located in the surface where the
densities are low. A projectile that is free of this undesirable
behavior is the photon.

In addition, the calculated strong differences in the in-medium
properties of longitudinal and transverse vector mesons can probably
only be verified in photon (real or virtual) induced reactions, where the
incoming
polarization can be controlled. Another approach
would be to measure the coherent photoproduction of vector mesons; here
the first calculation available so far\cite{PE} shows a distinct
difference in the production cross sections of transverse and
longitudinal vector mesons.

\subsection{Dilepton Production}
We have therefore -- also in view of a corresponding proposal for such
an experiment at CEBAF\cite{Preed} -- performed calculations for the
dilepton yield expected from $\gamma + A$ collisions. In these calculations
we have used the same transporttheoretical method as above for the
pion-induced dilepton emission, but now employ a better description
of the population of broad resonances\cite{BratEffe}.

\begin{figure}
  \centerline{\includegraphics[height=9cm]{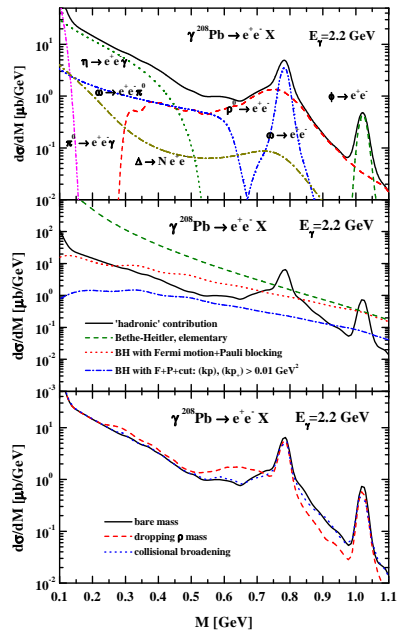}}
  \caption{Invariant mass spectra of dileptons produced in $\gamma +
  \mbox{}^{208}Pb$ reactions at 2.2 GeV. The top figure shows the
  various radiation sources, the middle figure the total yield from the
  top together with the Bethe-Heitler contributions, and the bottom part
  shows the expected in-medium effects (from \protect\cite{BratEffe}).}\label{gammadil}
\end{figure}

Results of these calculations are shown in Fig. \ref{gammadil}.
In the top figure the various sources of
dilepton radiation are shown. The dominant sources are again the same
as those in pion- and heavy-ion induced reactions, but the (uncertain)
$\pi N$ bremsstrahlung does not contribute in this reaction. The middle
part of this figure shows both the Bethe-Heitler (BH)
contribution and the contribution from all the hadronic sources. In the
lowest (dot-dashed) curve we have chosen a cut on the product of the
four-momenta of incoming photon ($k$) and lepton ($p$) in order to
suppress the BH contribution.
It is seen that even without BH subtraction the vector meson signal
surpasses that of the BH process.

The lowest
figure, finally, shows the expected in-medium effects\cite{BratEffe}: the
sensitivity in the region between about 300 and 700 MeV amounts to a
factor of about 3 and is thus in the same order of magnitude as in the
ultrarelativistic heavy-ion collisions.

\subsection{Photoabsorption}

Earlier in this paper I have discussed that a strong change of the
$\rho$ meson properties comes about because of its coupling to $N^* h$
excitations and that this coupling -- through a higher-order effect --
in particular leads to a very strong increase of the $\rho$ decay width
of the $N(1520) \: D_{13}$ resonance.

This increase may provide a reason for the observed disappearance of
the higher nucleon resonances in the photoabsorption cross sections on
nuclei\cite{Bianchi}. The
disappearance in the third region is a consequence of
the Fermi-motion. The disapparance of the second resonance
region, i.e. in particular of the $N(1520)$ resonance, however, presents
an interesting problem; it is obviously a typical in-medium effect.

\begin{figure}
  \centerline{\includegraphics[height=6cm]{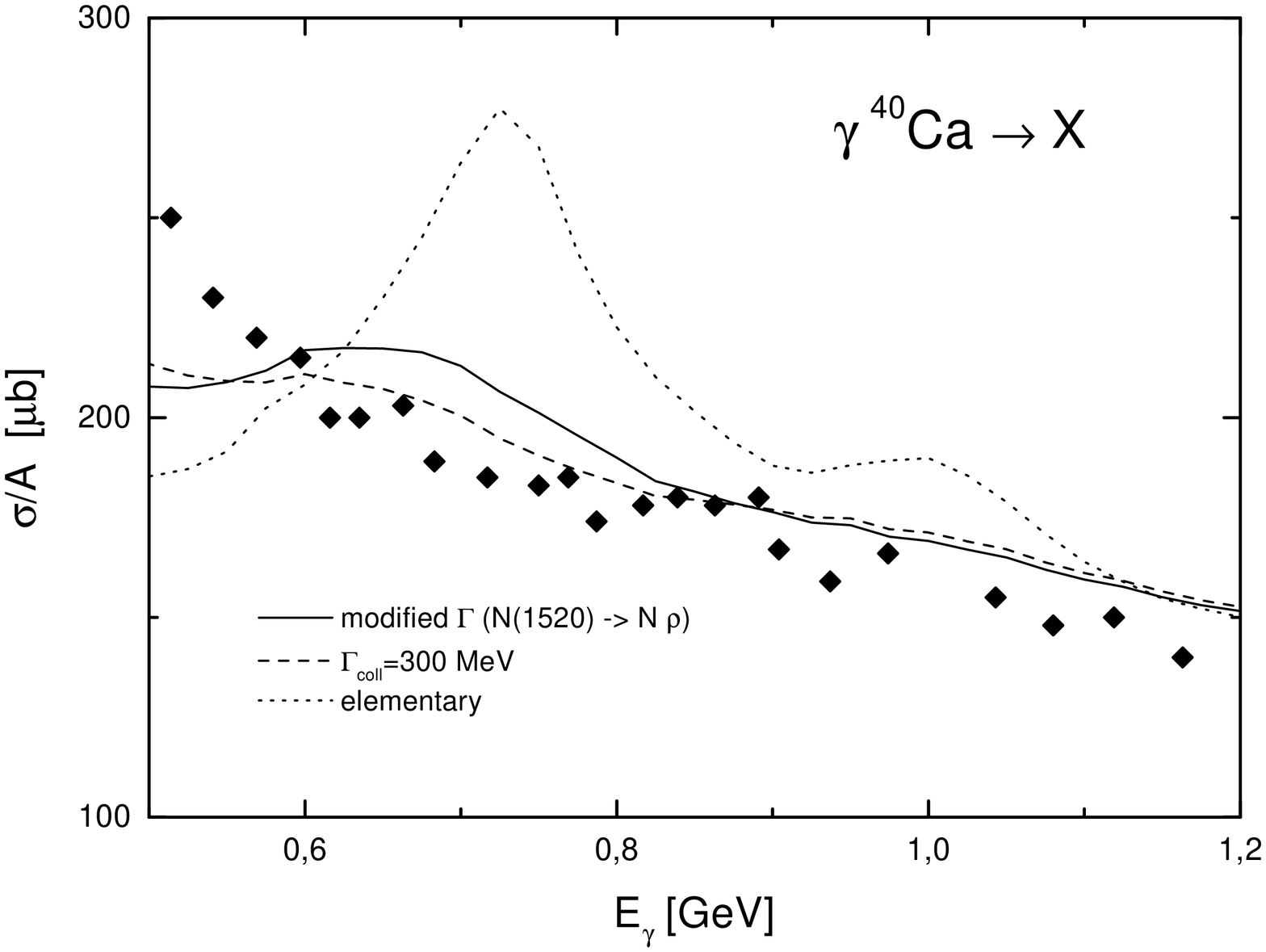}}
  \caption{Photoabsorption cross section in the second resonance region.
  Shown are the data from ref.\protect\cite{Bianchi}, the free
  absorption cross section on the proton, a Breit-Wigner fit with a total
  width of 300 MeV (dashed curve) and the result of a
  transporttheoretical calculation\protect\cite{Effeabs} with a medium
  broadened $\rho$ decay width of the $N(1520)$(solid curve).}\label{photabs}
\end{figure}

First explanations\cite{Kon} assumed a very strong collisional
broadening, but in ref.\cite{Effeabs} it has been shown that this
broadening is not strong enough to explain the observed disappearance
of the $D_{13}$ resonance. Since at the energy around 1500 MeV also the
$2 \pi$ channel opens it is natural to look for a possible connection
with the $\rho$ properties in medium. Fig.\ref{photabs}
shows the results of such an analysis (see also\cite{PE}). It is
clear that the opening of the phase space for $\rho$ decay of this
resonance provides enough broadening to explain its disappearance.
The talk by N. Bianchi at this conference\cite{Bianchitalk} covers this
topic in much more detail.

\section{Rho-meson properties at higher energies}

The calculations of spectral functions of $\rho$-meson properties in
medium so far have focussed on properties of rather slow mesons.
However, the relations (\ref{masswidth}) can be used to obtain some
information on the in-medium changes to be expected, because the cross
sections entering here contain all the possible couplings.

\begin{figure}
  \centerline{\includegraphics[height=6cm]{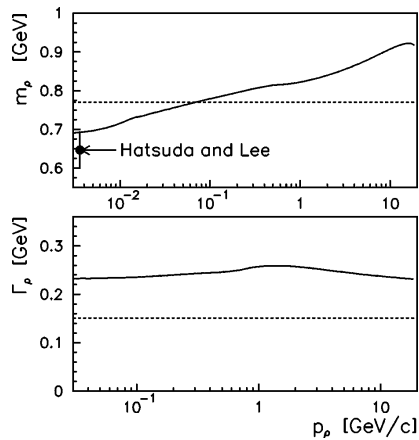}}
  \caption{$\rho$ meson mass and width as a function of momentum (from
  \protect\cite{KondrCass}).}\label{mgam}
\end{figure}

A first attempt in this direction was made by Eletsky and
Ioffe\cite{EJ} who obtained a \emph{positive} mass shift for $\rho$
mesons with momenta in the GeV region. This analysis has been linked to
the low-energy behavior by including all the relevant nucleon
resonances with $\rho$ meson decay branches by Kondratyuk et
al.\cite{KondrCass}. These authors find indeed that at momenta of about
100 MeV/c the mass shift according to (\ref{masswidth}) turns from
negative to positive; the width stays remarkably constant as a function
of $\rho$ momentum at a value of about 250 MeV, corresponding to a
collisional broadening width $\delta \Gamma \approx 100$ MeV (see
Fig.\ref{mgam}).

This number allows an easy estimate for the onset of shadowing in
high-energy photon-induced particle production experiments. For
example, in high-energy vector meson production\cite{Oneill} a crucial
scale is given by the so-called coherence length
\begin{equation}\label{coh}
  l_{\rm c} = \frac{2 \nu }{M_{\rm v}^2 + Q^2}
\end{equation}
where $Q$ is the four-momentum transfer, $\nu$ the energy transfer and
$M_{\rm v}$ the vector meson mass.
This length gives a measure for the distance over which the photon
appears as a rho meson. An obvious condition for the onset of shadowing
is then
$
  l_{\rm c} \ge \lambda_{\rho}
$
where $\lambda_\rho$ is the mean free path of the $\rho$ meson in
nuclear matter. The value of $\delta \Gamma_\rho$ of about 100 MeV
calculated in ref.\cite{KondrCass} translates into a value of
$\lambda_\rho$ of about 2 fm at high momenta. Using this value in
yields for the `shadowing threshold'
\begin{equation}\label{shad1}
\frac{2 \nu}{M_{\rm v}^2 + Q^2} \ge 2 \mbox{fm} ~.
\end{equation}
For example, at an energy transfer of 10 GeV, corresponding to the
Hermes regime, we obtain as condition for the shadowing regime
\begin{equation}\label{shadcond}
  Q^2 \le 1.5 \mbox{GeV}^2   ~.
\end{equation}
Increasing $Q^2$ (at fixed $\nu$) beyond this value yields a larger
transparency of the photon. This effect, which also shows up in the
calculations of Kopeliovich and Huefner\cite{KH}, must clearly be taken
into account in experiments of this sort searching for color
transparency; for further discussions of this exciting topic I refer to
the talk of T. O'Neill at this conference\cite{Oneill}.

\section{Summary}
In this talk I have concentrated on a discussion of the in-medium
properties of the vector mesons. I have shown that the scattering
amplitudes of these mesons on nucleons determine their in-medium
spectral functions, at least in lowest order in the
nuclear density. The dilepton channel can give information on the
properties of the $\rho$ and $\omega$ deep inside nuclear matter
whereas the $\pi$ decay channels --
because of their strong final state interactions -- can give only
information about the vector mesons in the low-density region.

The original QCD sum rule predictions of a lowered
vector meson mass have turned out to be too naive, because they were based on
the assumption of a sharp resonance state. In contrast, all specific
hadronic models yield very broad spectral functions for the $\rho$
meson with a distinctly different behavior of longitudinal and transvese
$\rho$'s. Recent QCD sum rule analyses indeed do not predict a lowering
of the mass, but only yield --
rather wide -- constraints on the mass and width of the vector mesons.
I have also discussed that hadronic models that include the coupling of
the $\rho$ meson to nucleon resonances and a corresponding shift of
vector meson strength to lower masses give a natural backreaction on
the width of these resonances. In particular, the $N(1520) D_{13}$
resonance is significantly broadened because of its very large coupling
constant to the $\rho N$ channel. Since the $\rho$ decay
of this resonance is
deduced only from a partial wave analysis of $2 \pi$ production\cite{Manley},
it would be very essential to have new, better data (and a new analysis of
these) for this channel.

A large part of the unexpected surplus of dileptons
produced in ultrarelativistic heavy-ion collisions can
presently be reproduced only by including a shift of $\rho$ strength to
lower masses. However, heavy-ion reactions are quite insensitive to
details of in-medium changes.

Motivated by this observation I have then discussed predictions for
experiments using pion and photon beams as incoming particles. In both
cases the in-medium sensitivity of the dilepton mass spectra is
comparable to that in heavy-ion experiments. In addition, such experiments
take place much closer to equilibrium, an assumption on which all
predictions of in-medium properties are based. Furthermore, only in such
experiments it will be possible to look for polarization effects. I have
also discussed the intriguing suggestion that the observed disappearance
of the second resonance region in the
photoabsorption cross section is due to the broadening of the $N(1520)$
resonance caused by the shift of $\rho$ strength to lower masses.

At high energies, finally, the in-medium broadening of the $\rho$ leads
to a mean free path of about 2 fm in nuclear matter. Shadowing will
thus only be essential if the coherence length is larger than this mean
free path. Increasing the four-momentum transfer $Q^2$ at fixed energy
transfer $\nu$ leads to smaller coherence length and thus leads to a larger
transparency. This effect is essential to verify experimentally; it is
superimposed on the color-transparency effect that is still being
looked for.

\end{document}